\documentclass[a4paper,10pt]{article}
\usepackage{graphics}
\usepackage{amssymb}
\begin{document}

\title{Behaviour of three charged particles on a plane under perpendicular magnetic field}
\author{A.Ralko and T.T.Truong, \\ 
Laboratoire de Physique Th\'eorique et Mod\'elisation,\\
 CNRS-Universit\'e de Cergy-Pontoise (UMR 8089). \\ 
F-95031, Cergy-Pontoise Cedex, France}

\maketitle

\begin{abstract}
We consider the problem of three identical charged particles on a plane under a perpendicular magnetic field and interacting through Coulomb repulsion. This problem is treated within Taut's framework, in the limit of vanishing center of mass vector $\vec{R} \to \vec{0}$, which corresponds to the strong magnetic field limit, occuring for example in the Fractional Quantum Hall Effect. 
Using the solutions of the biconfluent Heun equation, we compute the eigenstates and show that there is two sets of solutions. The first one corresponds to a system of three independent anyons which have their angular momenta fixed by the value of the magnetic field and specified by a dimensionless parameter $C \simeq \frac{l_B}{l_0}$, the ratio of $l_B$, the magnetic length, over $l_0$, the Bohr radius. This anyonic character, consistent with quantum mechanics of identical particles in two dimensions, is induced by competing physical forces. The second one corresponds to the case of the Landau problem when $C \to 0$. Finally we compare these states with the quantum Hall states and find that the Laughlin wave functions are special cases of our solutions under certains conditions.  \\ 
\emph{Pacs}:03.65.Ge, 05.30.Pr, 71.10.Pm, 73.43.-f
\end{abstract}

\section{Introduction}

Systems of interacting particles on a plane under a perpendicular magnetic field have been extensively studied, in particular in relation to the Fractional Quantum Hall Effect (FQHE) \cite{laughlin1}-\cite{truong}. Various kinds of wave functions (WF) have been proposed, such as the Laughlin WF or the composite fermions (CF) WF. The Laughlin WF \cite{laughlin1} is constructed from the one-particle states of the Landau problem in the lowest Landau level (LLL) and generalized to $N$ particles with the help of a Jastrow factor. In another way, the CF WF, proposed by Jain \cite{jain1}-\cite{jain3}, starts with the $N$-body WF of $N$ planar electrons in uniform magnetic field to which $2m$ vortices of unit quantum flux $\phi_0$ are attached. Jain has chosen this ansatz so that this WF, projected onto the LLL corresponds to the WF of $N$ interacting particles.  These eigenstates describe to a good approximation the main characteristics of the FQHE, but they are not analytical solutions of the Schr\"odinger equation.

Recently, exact solutions of few-particle systems have been investigated, i.e. two or three, and it shows that these states contain the quantum Hall states \cite{taut1,truong}. The case of two particles has been also considered from different points of view. Taut \cite{taut1} has treated this problem by adding an oscillator of frequency $\omega_0$, which has its values fixed by the angular momentum as $m$ is quantized in the standard way. Khare  \cite{khare} \& Ver\c{c}in \cite{vercin} have considered that the charged particles are anyons with an anyonic parameter $\alpha$, that the angular momentum takes integer values and, this forces the external magnetic field to have discrete values. Truong \& Bazzali \cite{truong} consider two spinless electrons and use the fact that the relative motion of two identical planar particles can have an anyonic behavior  \cite{laidlaw}-\cite{khurana}, characterized by fractional values of the angular momentum, as specified by quantum mechanics of identical planar particles. Here, the angular momentum values are fixed by the magnetic field and determine the expression of the eigenstates.  

For three particles, the problem is more difficult and cannot be solved analytically in the standard approach. Recently, Taut  \cite{taut2} has proposed an interesting change of variables which leads to analytical solutions of the three body problem in the $R$$\to 0$ limit. This limit can be obtained in the FQHE scheme, where the magnetic field is very strong. But we show that the results remain very accurate beyond this limit. We find that the WF of the system is the product of three WF's of one quasi-particle in magnetic field and rescaled Coulomb repulsion. The anyonic character is introduced in a natural way by the competition of the magnetic confinement and the Coulomb repulsion. The interesting fact is that our solutions are always eigenstates of the Schr\"odinger equation for all values of the magnetic field which is fixed by hand from the start. 

This anyon picture has an important mathematical and physical relevance. Objects appear in a natural way when we consider the relative motion of identical particles and tow dimensional space. The group governing the exchange of identical objects is not the standard permutation group, but the Braid group \cite{khare,wilczek,lerda}, allows us to consider all possibilities between fermion statistics and boson statistics. These new particles arise with fractional statistics and are called anyons. Physically, in FQHE experiments, anyonic behavior is formed in excitations. Effectively, when we consider the first excited state of the Laughlin ground state, one can see that these excitations are quasi-particles with fractional quatum numbers, with fractional charge $\frac{e}{3}$ for the quasi-hole \cite{laughlin2,chakraborty,prange}. So it seems natural to consider, for a system of identical spinless particles in two dimensions, anyonic states as possibility of solutions.  

For the rest of the paper, we present in section 2 the Taut coordinate transformation and discuss the $R \to 0$ limit. We transform the so-obtained equation and expand in a multi-pole series the Coulomb interaction terms. In section 3, we solve the equation using biconfluent Heun functions, give the energy spectrum of the system and discuss the conditions of existence of these solutions. In section 4, we present the explicit expressions of the three first levels and describe the general features for level $p$. We improve, in section 5, the accuracy of the limit $R \to 0$ by calculating the energy contribution of the second and third term of the multi-pole expansion. Finally, we give some relations, in section 6, with the two-particle solution and compare our WF with the quantum Hall states (Laughlin). Some remarks and conclusions are given in the last section 7. 

\section{Transformation of the Schr\"odinger equation}

The Hamiltonian of three electrons of charge $-e$ and mass $m^{*}$ on a plane under a perpendicular magnetic field $B$ and interacting via the Coulomb repulsion reads:
\begin{eqnarray}
H = \sum_{j=1}^{3} \frac{- \hbar^2}{2m^{*}}\left[\vec{\nabla}_j +\frac{ie}{\hbar} \vec{A}_j \right]^2 +  \sum_{j<k} \frac{\kappa e^2}{|\vec{r_j} - \vec{r_k}|}
\label{eqn:01}
\end{eqnarray}
where $\kappa = (4 \pi \epsilon_0)^{-1}$ in S.I. units and $\vec{A}_j = \frac{B}{2} \left( y_j \vec{u}_x - x_j \vec{u}_y  \right)$ in the symmetric gauge. This yields a constant magnetic field $\vec{B} = - B \vec{u}_z$. $\vec{r_j}$ is the position of the $j$th particle. 

Following Taut \cite{taut2} , we seek a change of variables which leaves the kinetic part of the Hamiltonian unchanged and transforms the interaction terms in such a way to have only a dependence on one of the new coordinates and the center-of-mass. An orthogonal transformation fullfilling these conditions may be obtained with the choice:
\begin{eqnarray}
\left[ \begin{array}{c}\vec{\xi}_{1} \\ \vec{\xi}_{2} \\ \vec{\xi}_{3} \end{array} \right] =\left[ \begin{array}{ccc}\frac{1}{3} & a & b \\ b & \frac{1}{3} & a \\ a & b &  \frac{1}{3} \end{array} \right] \left[ \begin{array}{c}\vec{r}_{1} \\ \vec{r}_{2} \\ \vec{r}_{3} \end{array} \right]
\label{eqn:02}
\end{eqnarray}
where $a= 1/3 - 1/\sqrt{3}$ and $b= 1/3 + 1/\sqrt{3}$. In this change of variables the relative distances between any two of the three particles are now:
\begin{eqnarray}
\label{eqn:03}
|\vec{r}_1 -\vec{r}_2| = \sqrt{3} |\vec{\Xi}- \vec{\xi}_3| \nonumber \\
|\vec{r}_2 -\vec{r}_3| = \sqrt{3} |\vec{\Xi}- \vec{\xi}_1| \\
|\vec{r}_3 -\vec{r}_1| = \sqrt{3} |\vec{\Xi}- \vec{\xi}_2|, \nonumber
\end{eqnarray}
where $\vec{\Xi} = \frac{1}{3}(\vec{r}_1+\vec{r}_2+\vec{r}_3)$ and so $\vec{R}=\vec{\Xi}$.
As the Coulomb repulsion between the particles has rotational symmetry, we use cylindrical coordinates to describe the magnetic field. The Hamiltonian in the new coordinates is thus:
\begin{eqnarray}
H = \sum_{j=1}^{3} \left\{ \frac{- \hbar^2}{2m^{*}}\left[\vec{\nabla}_j -\frac{ieB}{2\hbar}\xi_j \vec{u}_{\theta_j}  \right]^2 +  \frac{\kappa e^2}{\sqrt{3} |\vec{\Xi} - \vec{\xi}_j|} \right\}
\label{eqn:04}
\end{eqnarray}

To discuss the center-of-mass motion, we introduce the Jacobi variables:
\begin{eqnarray}
\vec{R}= \frac{\vec{r}_1 +\vec{r}_2+\vec{r}_3}{3}, \ \ \ \vec{\eta}=\frac{\vec{r}_2 -\vec{r}_1}{\sqrt{2}}, \ \ \ \ \vec{\zeta}=\sqrt{\frac{2}{3}}\left( \vec{r}_3-\frac{\vec{r}_1 +\vec{r}_2}{2} \right).
\label{eqn:05}
\end{eqnarray}

The Hamiltonian of the center-of-mass is then:
\begin{eqnarray}
H_{CM} = \frac{-\hbar^2}{2 M}\left(\vec{\nabla}_{R} + \frac{i e_{R} B}{2 \hbar} R \vec{u}_{R} \right)^2.
\label{eqn:06}
\end{eqnarray}
It describes the motion of a planar particle of charge $e_R = -3 e$ and mass $M = 3 m$ in uniform perpendicular magnetic field $B$ . The solution of the Schr\"odinger equation is well known and the semi-classical picture is a circular orbit with a radius inversely proportional to $B$. So the high field limit forces $|\vec{R}|=R\to 0$ and we shall consider this regime. For the rest of the paper, $R$ is assumed to be smaller than the average distance of the electron separation $\xi_i$.

Let us go back to the Taut's variables. As it was shown, $\vec{R} = \vec{\Xi}$, so the quasi-particles have their center-of-mass vector $\vec{\Xi} \to \vec{0}$ in our limit. We can now expand the interaction terms of Eq.(\ref{eqn:04}) in a multi-pole series:
\begin{eqnarray}
\sum_{i=1}^{3} \frac{1}{|\vec{\xi}_i - \vec{\Xi}|} = \sum_{i=1}^{3} \frac{1}{\xi_i} + \sum_{i=1}^{3} \frac{\vec{\Xi}.\vec{\xi_i}}{\xi_{i}^{3}} + \frac{1}{2} \sum_{i=1}^{3} \left( \frac{3(\vec{\Xi}.\vec{\xi_i})^2}{\xi_{i}^{5}} - \frac{\Xi^2}{\xi_{i}^{3}} \right) + ...
\label{eqn:07}
\end{eqnarray}

So Eq.(\ref{eqn:04}), reads to the lowest order:
\begin{eqnarray}
H = \sum_{j=1}^{3} \left[ \frac{- \hbar^2}{2m^{*}}\left[\vec{\nabla}_j -\frac{ieB}{2\hbar}\xi_j \vec{u}_{\theta_j}  \right]^2 +  \frac{\kappa e^2}{\sqrt{3} \xi_j} \right]
\label{eqn:08}
\end{eqnarray}
 and it is mathematically equivalent to setting directly $|\vec{\Xi}|=\Xi=0$ in Eq.(\ref{eqn:04}). As Taut says \cite{taut2}, it does not mean that the coordinates $\xi_i$ are no longer independent, but only that we just keep the zeroth order term in the multi-pole expansion of the interaction term. The one-particle part of the Hamiltonian is independent of $\Xi$ and stay exact. 

It is the Hamiltonian of three quasi-particles in magnetic field, interacting via an electrostatic field of a central charge $Q = -e/\sqrt{3}$. Since these  quasi-particles are independent, the Schr\"odinger equation,
\begin{eqnarray}
H \Psi(\xi_1,\theta_1,\xi_2,\theta_2,\xi_3,\theta_3) = E \Psi(\xi_1,\theta_1,\xi_2,\theta_2,\xi_3,\theta_3)
\label{eqn:09}
\end{eqnarray}
appears as a sum of three independent one body Schr\"odinger equation:
\begin{eqnarray}
&&\left(\sum_{j=1}^{3} h_j \right) \phi(\xi_1,\theta_1)\phi(\xi_2,\theta_2)\phi(\xi_3,\theta_3) = \nonumber \\
&&\left(\sum_{j=1}^{3} E_j \right) \phi(\xi_1,\theta_1)\phi(\xi_2,\theta_2)\phi(\xi_3,\theta_3)
\label{eqn:10}
\end{eqnarray}
where the eigen energy is a sum $E = E_1 + E_2 + E_3$. Thus $\Psi$ is just the product of each WF of the quasi-particles $ \Psi(\xi_1,\theta_1,\xi_2,\theta_2,\xi_3,\theta_3)= \phi(\xi_1,\theta_1)\phi(\xi_2,\theta_2)\phi(\xi_3,\theta_3) $. 
We are now reduced to solve the sub-Schr\"odinger equation for one quasi-particle defined by:
\begin{eqnarray}
\left\{ \frac{- \hbar^2}{2m^{*}}\left[\vec{\nabla} -\frac{ieB}{2\hbar}\xi \vec{u}_{\theta}  \right]^2 +  \frac{\kappa e^2}{\sqrt{3} \xi} -E \right\} \phi(\xi,\theta)=0
\label{eqn:11}
\end{eqnarray}

Note that since we are in the zeroth order limit, relations (\ref{eqn:03}) reads:
\begin{eqnarray}
 \xi_1 = |\vec{r_2} -\vec{r_3}|/\sqrt{3} \nonumber \\
 \xi_2 = |\vec{r_3} -\vec{r_1}|/\sqrt{3} \\
 \xi_3 = |\vec{r_1} -\vec{r_2}|/\sqrt{3} \nonumber
\label{eqn:12}
\end{eqnarray}
 so Eq.(\ref{eqn:11}) has the same form as in [6] and the problem is nothing but the quantized relative motion of two spinless electrons. We are exactly in the conditions where  anyonic statistics can occur, and analytic solutions exist.
\\

\emph{Remarks:} Since we deal with two dimensional identical particles, the anyonic solution must exist even if we introduce the terms of order $0(R)$. However, the form of the WF must change because these terms depend on $R$ and $\xi_i$, so they modify the statistics parameter and the solutions. We cannot compute exactly these solutions, but as we will see, our approximation is valid in a more general scheme than the strong magnetic field scheme. In this case, as it shows in section $5$, we can keep our solutions when we re-introduce the $0(R)$ terms since their contributions are very small.

\section{Construction of the solutions}

Using the magnetic length $l_{B}^{2} = \frac{\hbar}{e B}$, the Bohr radius $l_0 = \frac{\hbar^2}{\kappa e^2 m^{*}}$ and the rescaled energy $\epsilon = \frac{2 m^{*} E l_{B}^{2}}{\hbar^2}$, we rewrite Eq.(\ref{eqn:11}) in the form:
\begin{eqnarray}
\left\{ \frac{1}{\xi} \frac{\partial}{\partial \xi} \xi \frac{\partial}{\partial \xi} +\frac{1}{\xi^2} \frac{\partial^2}{\partial \theta^2} - \frac{\xi^2}{4 l_{B}^{2}} - \frac{2}{\sqrt{3} l_0 \xi} + \frac{\epsilon}{l_{B}^{2}}-i\frac{1}{l_{B}^{2}}\frac{\partial}{\partial \theta} \right\} \phi(\xi,\theta)=0
\label{eqn:13}
\end{eqnarray}
where $\xi$ is the particle separation and $\theta$ its polar angle with respect to a reference axis. It is important to note that we do not assume from the outset that the particles are anyons, thus there are no statistical potentials presents as in \cite{khare}. Since $\theta$ is a cyclic variable, it is clear that $p_{\theta} = \frac{\hbar}{i}\frac{\partial}{\partial \theta}$ is a conserved quantity. Moreover, as shown by Leinaas and Myrheim \cite{leinaas}, the momentum $p_{\theta}$ for a system of identical particles in two dimensions has continuous spectrum. Thus we can make the standard separation of variables under the form:
\begin{eqnarray}
\phi(\xi,\theta) = \frac{e^{im\theta}}{\sqrt{2 \pi}} F(\xi)
\label{eqn:14}
\end{eqnarray}
with now $m \in \mathbb{R}$. Note that $m$ is the angular momentum value of one of the quasi-particles and in the total WF to differenciate the three different possible values, we set $m_{ij}$ as relative angular momenta of the particles $i$ and $j$, i.e. the corresponding quasi-particle coordinate $\xi_j = |\vec{r_i}-\vec{r_j}|/\sqrt{3}$. In its dimensionless form, Eq.(\ref{eqn:13}), with $\xi = \sqrt{2} l_B x$ and $C = \frac{2 \sqrt{2} l_B}{\sqrt{3}l_0}$, reads:
\begin{eqnarray}
F''(x) +\frac{1}{x}F'(x)+ \left\{2(\epsilon+m)-\frac{C}{x}-x^2-\frac{m^2}{x^2}\right\}F(x)=0 .
\label{eqn:15}
\end{eqnarray}

From this equation, it can be seen that the asymptotic form of $F(x)$ is $e^{-\frac{x^2}{2}}$ \cite{ballentine} when $x \to \infty$, so we make the change of function $F(x) = u(x)e^{-\frac{x^2}{2}}   $. The equation in $u(x)$ is now:
\begin{eqnarray}
x u''(x) +(1-2x^2)u'(x)- \left\{C-2(\epsilon+m-1)+\frac{m^2}{x}\right\}u(x)=0 .
\label{eqn:16}
\end{eqnarray}

Near the origin, this equation takes the limiting form:
\begin{eqnarray}
u_{as}''(x) + \frac{u_{as}'(x)}{x}- \frac{m^2}{x^2}u_{as}(x)=0
\label{eqn:17}
\end{eqnarray}
and its solution is $u_{as} = x^{s}$ where $s=\pm |m|$. Finally we further change  function $u(x)= x^{s} N(x)$, where $N(x)$ obeys now a biconfluent Heun equation of canonical form:
\begin{eqnarray}
x N'' + (1+\alpha - \beta x - 2 x^2) N' \nonumber \\
+ \left((\gamma-\alpha-2)x-\frac{1}{2}(\delta+\beta(1+\alpha)) \right)N=0.
\label{eqn:18}
\end{eqnarray} 
This equation generalizes the Kummer's equation, and depends on four parameters $\alpha,\beta,\gamma,\delta$ \cite{ronveaux}-\cite{batola} . In our case the parameters take the values: $\alpha = 2 s$, $\beta = 0$, $\gamma = 2 (\epsilon + m)$ and $\delta=2 C$. The biconfluent Heun function is explicitely given by the series:
\begin{eqnarray}
N(2s,0,2 (\epsilon + m),2C;x) = \sum_{\nu = 0}^{\infty} \frac{A_{\nu}}{(1+2s)_{\nu}}\frac{x^{\nu}}{\nu !},
\label{eqn:19}
\end{eqnarray} 
where the coefficients $A_{\nu}$ fulfills a three way recursion relation:
\begin{eqnarray}
A_{\nu+2} = C A_{\nu+1} - 2(\nu+1)(\nu+1+2s)(\epsilon+m-s-2-\nu)A_{\nu}
\label{eqn:20}
\end{eqnarray}
with $A_0 = 1$ and $A_1 = C$.

As $N(2s,0,2 (\epsilon + m),2C;x)$ grows as $e^{x^2}$ when $x \to \infty$ \cite{batola} , it is necessary to cut its defining power series down to a polynomial by imposing two conditions on the recursion relation (\ref{eqn:20}):
\begin{itemize}
\item energy quantization $\epsilon = \epsilon_n = n+1+s-m$ with $n=1,2,3,...$;
\item and $A_{n+1} = 0$.
\end{itemize}
The second condition yields some relations between the angular momentum $m$ and the ratio parameter $C$ in term of polynomials in $C^2$ of degree $\frac{n+1}{2}$ if $n$ is odd or $\frac{n}{2}$ if $n$ is even. This has been already pointed out in  \cite{truong}. The consequence is that the magnetic field, in competition with the Coulomb repulsion, fixes $m$ and thus gives rise to the expected anyonic behavior. The non-normalized WF is finally:
 \begin{eqnarray}
\phi(x,\theta) = \frac{e^{i m \theta}}{\sqrt{2 \pi}} x^{\pm |m|} \Pi_n(C,x)\exp \left(-\frac{x^2}{2}\right)
\label{eqn:21}
\end{eqnarray}
where $\Pi_n(x)$ is a special biconfluent Heun polynomial. Since $m \in \mathbb{R}$, there are singular cases where the WF diverges but remain square integrable, i.e. when $-1<-|m|<0$. For the rest of the paper, we restrict ourselves to the regular cases $s=+|m|$, the singular cases being already discussed in  \cite{ralko} for two particles. So the energy spectrum is minimum for positive values of the angular momentum $m > 0$. The normalization constant can be easily computed and is given in  \cite{truong}.

Note that in the case $C=0$, we have the correspondence with the Kummer's functions  \cite{ronveaux}:
\begin{eqnarray}
N(2s,0,2 (\epsilon + m),0;x) = \phi( -\frac{1}{2}(\epsilon+m-s-1),1+s;x^2),
\label{eqn:22}
\end{eqnarray}
and when $-\frac{1}{2}(\epsilon+m-s-1) = -k$, we have the Laguerre's polynomials:
\begin{eqnarray}
N(2s,0,2 (\epsilon + m),0;x) = k! \frac{\Gamma(1+|m|)}{\Gamma(1+|m|+k)} L_{k}^{|m|}(x^2).
\label{eqn:23}
\end{eqnarray}
 These solutions correspond to the Landau problem, with an energy spectrum $\epsilon_k = (2k + |m| - m +1)$ where $k=0,1,2,...$. So we have state degeneracy when the energies verify $\epsilon_n = \epsilon_k$, or when $n = 2k$. In other words, when $n$ is even, our solution fits exactely with the Landau states (Laguerre polynomials) and we have two solutions for two regimes of $B$. The quasi-particles can be anyons if $C$ is finite, or just electrons in a magnetic field if $C=0$,  infinite $B$ or no Coulomb terms. We can compute these two kinds of solutions, by choosing a new parameter $p=0,1,2,...$, and the energy is now given in each cases by $\epsilon_p =p + 1$ (for $m>0$). The case $p=0$ is the so-called LLL, and the possibility of Landau states occurs for $p$ even. In this case, the solution of Eq.(\ref{eqn:15}) is:
\begin{eqnarray}
\phi(x,\theta) = \frac{e^{i m \theta}}{\sqrt{2 \pi}} x^{|m|} \Pi_p(C,x) \exp \left(-\frac{x^2}{2} \right) 
\label{eqn:24}
\end{eqnarray} 
with   $\Pi_p(C,x)$:
\begin{itemize} 
\item if $p$ odd, the biconfluent Heun's polynomial;   
\item and if $p$ even,  $\Pi_p(0,x)$ corresponds to the Landau states (Laguerre's polynomials) whereas  $\Pi_p(C \neq 0,x)$ corresponds to the anyonic states.
\end{itemize}

\section{Some explicit expressions}

We present here the explicit forms of solutions for three levels, $p=0,1,2$, then we generalize the procedure to the case of $p$ even and odd. 
\\

\emph{$p=0$:}
\\
The rescaled eigen-energy is $\epsilon_0 = 1$ and the second condition of quantization give rises to $A_1 =0$. But we have seen from the recursion relation (\ref{eqn:20}) that $A_1 = C$. This shows that the solution is a limiting case of the biconfluent Heun equation and $C=0$ for the Landau states. Thus the solution of Eq.(\ref{eqn:18}) is a Laguerre polynomial for $p=0$:
\begin{eqnarray}
N(2|m|,0,2(1+|m|),0;x) = \Pi_0 (0,x) = L_{0}^{m}(x^2) = 1
\label{eqn:25}
\end{eqnarray} 

 The limit, $C=0$, corresponds to an infinite magnetic field or vanishing Coulomb interaction terms. There's no anyonic behavior, and the angular momentum's values are independent of $C$, so the quasi-particles are here standard particles (electrons). Thus we have a degeneracy in $m$ for a given $C$ and each of the quasi-particles can take a different value of $m$ called $m_{ij}$, the relative angular momentum between particles $i$ and $j$. It is important to note that when we have anyonic states, all $m_{ij}$ are equals since the values are fixed by the same relation in $C^2$ and when the states are asymptotic, one can have different or same $m_{ij}$.
\\

\emph{$p=1$:}
\\
It corresponds to $n=1$ but there is no connection with $k$. So we expect to have only anyonic states and no Landau states. To verify this hypothesis, we look at the second quantization condition $A_2 =0$:
\begin{eqnarray}
A_2 = 0 = C^2 -2(1 + 2|m|),
\label{eqn:26}
\end{eqnarray} 
with the solution $|m| = \frac{1}{2}\left(\frac{C^2}{2}-1 \right)$. This is the expected anyonic behavior, the angular momentum is fixed by the magnetic field's value via the ratio parameter $C$ in a natural way. The quasi-particle is here an anyon. The solution of  Eq.(\ref{eqn:18}) is:
\begin{eqnarray}
N(2|m|,0,2(1+|m|),2C;x) = \Pi_1 (C,x) = 1 + \frac{2}{C}x.
\label{eqn:27}
\end{eqnarray} 

\begin{figure}[ht]
\caption{\label{fig:01}Distribution of probability versus $r = \xi/l_0$ for $p=1$ and for $3$ values of $C^2=6,10$ and $14$ corresponding respectively to $m=1,2$ and $3$.}
\begin{center}
\resizebox{0.6\textwidth}{!}{
\includegraphics{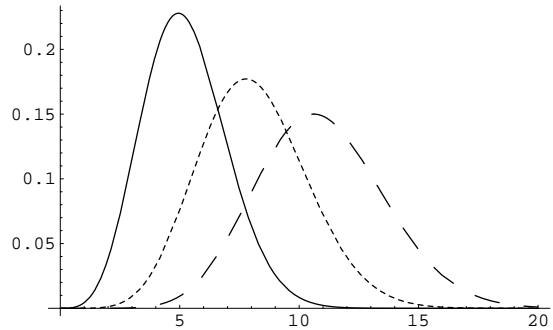}}
\end{center}
\end{figure}

Contrary to the case $p=0$, there's no Landau states here for two reasons. The eigen-energy value does not exist in this case and the second condition of quantization does not give this solution. The interesting aspect is that since $m$ is entirely fixed by $C$, all the quasi-particles have consistent $m_{ij}$ and it is a common point with the quantum Hall states. We have plotted in Fig.(\ref{fig:01}) the normalised distribution of probability
\begin{eqnarray}
P(C,r) = \frac{r |\phi(r,\theta)|^2}{<\phi(r,\theta)|\phi(r,\theta)> }
\label{eqn:28}
\end{eqnarray} 
 with the rescaled variable $r = \xi / l_0$ and for different values of $C$. This state has no nodes and corresponds to a ground state. We can see for larger $C$ one have larger probability. It is natural that when $B$ decreases, the particles are less confined and that the inter-particles distance increases.

\begin{figure}[ht]
\caption{\label{fig:02}Distribution of probability versus $r = \xi/l_0$ for $p=2$ and for $3$ values of $C^2=20,28$ and $36$ corresponding respectively to $m=1,2$ and $3$.}
\begin{center}
\resizebox{0.6\textwidth}{!}{
\includegraphics{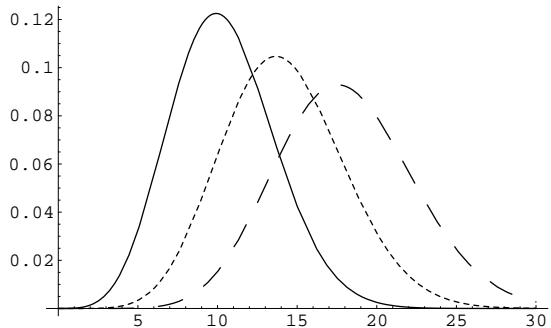}}
\end{center}
\end{figure}

\emph{$p=2$:}
\\
For the first time, there is a connection with $n$ and $k$. So we expect to find two kinds of solutions. Let us reconsider the second condition of quantization:
\begin{eqnarray}
A_3 = 0 = C \left(C^2 - 12 - 16 |m| \right)
\label{eqn:29}
\end{eqnarray} 
We have two solutions:
\begin{itemize}
\item $C=0$ this yields a Landau states for $k=1$ with:
\begin{eqnarray}
\Pi_2 (0,x) = L_{1}^{m}(x^2) =1 - \frac{x^2}{1+|m|},
\label{eqn:30}
\end{eqnarray}
\item $|m|=\frac{C^2-12}{16}$ which corresponds to an anyonic state for $n=2$ with:
\begin{eqnarray}
\Pi_2 (C,x) = 1 + \frac{C}{1+2|m|}x + \frac{C^2-4(1+2|m|)}{(1+2|m|)(2+2|m|)}\frac{x^2}{2}.
\label{eqn:31}
\end{eqnarray} 
\end{itemize}
Here, for the same value of the energy, we have two sets of solutions for a different range of $C$. The first set is the Landau states, at the first level $p=0$, and the other an anyonic states where the value of $m$ is defined by $C$. We present in Fig.(\ref{fig:02}) the distribution of probability for the anyonic states at $p=2$, i.e. solutions of the second condition of quantization with finite $C$. As it is expected, the maxima of probabilities are larger in $r$ than in the case $p=1$. Note that this state is an excited state without nodes, this is different from the asymptotic case $p=2$ and $C=0$, which has only one node. The behavior, for an increasing $C$ is the same as for $p=1$ and it agrees with the physics.  

Generally speaking, one can consider $2$ separate cases: $p$ odd and $p$ even. When $p$ is odd, the condition $A_{p+1}=0$ yields a relation between $m$ and $C^2$, which is a polynomial in $C^2$ of degree $\frac{p+1}{2}$. So we have $\left( \frac{p+1}{2}\right)$ solutions for $m$ and each corresponds to an anyonic state for which the distribution of probability has $1$ to $\frac{p+1}{2}$ nodes.

When $p$ is even, the degree of the polynomial is $\frac{p}{2}$ and the condition is of type: $C\left\{ \textrm{polynomial of degree} \ \frac{p}{2} \right\}$. Thus we have the anyonic solution in the same way as for $p$ odd, plus an overall factor $C$ which leads to the $C=0$ states, i.e. the Landau states. Hence we have shown the existence of two sets of solutions depending of the value of $C$. It is interesting to note that analytic solutions can be obtained only for $p < 9$ since polynomials of higher order than $4$ do not have closed form expressions for their zeros. The anyonic states have a number of nodes between $1$ and $\frac{p}{2}$. The number of nodes of the asymptotic states is equal to $\frac{p}{2}$.

The main difference between the two sets of solutions is the interaction dependence. For the anyonic solutions, $m$ is defined by $C$ which represents the competition of the two forces present in the problem. Thus going from weak to strong coupling, the states change via $m$, showing the strong sensitivity to the interaction term. Conversely, for the $C=0$ states, we have seen that $m$ is independent of $C$. This shows a strong stability of the solutions under the interaction.

Finally, to give an idea of the magnetic field value under which the particles evolve, for $p=1$, in GaAs $C^2 \simeq \frac{76.2119}{B(T)}$ and when $m=3$ we find a magnetic field of $B \simeq 5.44 \  T$, which it is a characteristic value in the FQHE.

\section{Validity of the approximation}

We have seen that our limit $R \to 0$ is equivalent to put $\vec{\Xi} = \vec{0}$ in Eq.(\ref{eqn:04}) and that is does not mean that the coordinates $\xi_i$ are no longer independent, but that we just keep the zeroth order term in the multi-pole expansion. In this section, we want to estimate the error to see whether by neglecting the dipole and the quadrupole term in the expansion (\ref{eqn:07}) or not, this treatment is also acceptable in a more general way than in the high magnetic field limit.
\\

\emph{Dipole and quadrupole energy contribution}
\\
It is easy to prove, for one anyonic quasi-particle and for states with equals $m_{ij}$, that the first order term in the expansion of Coulomb interaction, i.e. the dipole contribution is:
\begin{eqnarray}
\epsilon^{(1)} = \frac{C}{6} I^{(-1)}(p,m),
\label{eqn:32}
\end{eqnarray} 
the next quadrupole contribution is:
\begin{eqnarray}
\epsilon^{(2)} = \frac{C}{18} \left( I^{(-1)}(p,m) - I^{(2)}(p,m) I^{(-3)}(p,m) \right),
\label{eqn:33}
\end{eqnarray} 
where we have defined
\begin{eqnarray}
I^{(\alpha)}(p,m) &=& N_{x}^{2} <\phi(x,\theta)|x^{\alpha} |\phi(x,\theta)> \nonumber \\
&=& N_{x}^{2} \int_{0}^{\infty} x^{2|m|+1+\alpha} \left[\Pi_p(C,x) \right]^2  \exp \left(-x^2\right) dx,
\label{eqn:34}
\end{eqnarray} 
with the normalization constant
\begin{eqnarray}
N_{x}^{-2} = \int_{0}^{\infty}  x^{2|m|+1} \left[\Pi_p(C,x) \right]^2  \exp \left(-x^2\right) dx.
\label{eqn:35}
\end{eqnarray} 
\\

\emph{Results and comments}
\\
In tables [1] and [2], we have compared these values with the zero-order energy $\epsilon_{p}^{0} =( p +1+|m|-m)$ for which we have choosen $s=+|m|$.
\begin{table}[ht]
\begin{center}
\caption{Energy calculations for $p=1$ and different values of $m$ with the corresponding magnetic field. $\epsilon_{+}^{(0)}$ is for $m>0$ and  $\epsilon_{-}^{(0)}$ is for $m<0$. The error is calculated by summing all the energies for $m>0$ and comparing the total with  $\epsilon_{+}^{(0)}$. The case $|m| \to \infty$ is a limiting case evaluated from the results obtained with great values of $C^2$.}
\begin{tabular}{@{}llllllll}
$|m|$ & $C^2$ & $B$ & $\epsilon_{-}^{(0)}$ & $\epsilon_{+}^{(0)}$ & $\epsilon^{(1)}$ & $\epsilon^{(2)}$ & $\%$ \\
$\frac{3}{4}$ & $5$ & $15.24$ & $3.5$ & $2$ & $0.30438$ & $-0.33895$ & $1.73$ \\
$1$ & $6$ & $12.70$ & $4$ & $2$ & $0.30866$ & $-0.20342$ & $5.25$ \\
$2$ & $10$ & $7.62$ & $6$ & $2$ & $0.31783$ & $-0.08706$ & $11.53$ \\
$3$ & $14$ & $5.44$ & $8$ & $2$ & $0.32203$ & $-0.05663$ & $13.27$ \\
$4$ & $18$ & $4.23$ & $10$ & $2$ & $0.32444$ & $-0.04211$ & $14.16$ \\
$5$ & $22$ & $3.46$ &  $12$ & $2$ & $0.32601$ & $-0.03356$ & $14.62$ \\
$10$ & $42$ & $1.81$ & $22$ & $2$ & $0.32943$ & $-0.01669$ & $15.6$ \\
$20$ & $82$ & $0.93$ & $42$ & $2$ & $0.33132$ & $-0.00833$ & $16.14$ \\
$50$ & $202$ & $0.38$ & $102$ & $2$ & $0.33251$ & $-0.00333$ & $16.46$ \\
$100$ & $402$ & $0.19$ & $202$ & $2$ & $0.33292$ & $-0.00166$ & $16.56$ \\
$\infty$ & $\infty$ & $0$ & $\infty$ & $2$ & $\frac{1}{3}$ & $0$ & $\frac{100}{6}$ \\
\end{tabular}
\end{center}
\end{table}
\begin{table}[ht]
\begin{center}
\caption{Calculation of total energies in different levels. The error is calculated in the same way as in Table[1]. We have choosen the value of $m$ fixed at $1$.}
\begin{tabular}{@{}lllllll}
$p$ & $C^2$ & $B$ & $\epsilon_{+}^{(0)}$ & $\epsilon^{(1)}$ & $\epsilon^{(2)}$ & $\%$ \\
$1$ & $6$ & $12.70$ & $2$ & $0.30866$ & $-0.20342$ & $5.25$ \\
$2$ & $28$ & $2.72$ & $3$ & $0.58074$ & $-0.25351$ & $10.90$ \\
$3$ & $72.55$ & $1.05$ & $4$ & $0.83210$ & $-0.25483$ & $14.43$ \\
\end{tabular}
\end{center}
\end{table}
\begin{figure}[ht]
\caption{\label{fig:03} Error on the energy of states with $m >0$ versus magnetic field B(T) in the $p=1$ level. The point at $B=0$ is an extrapolation.  }
\begin{center}
\resizebox{0.6\textwidth}{!}{
\includegraphics{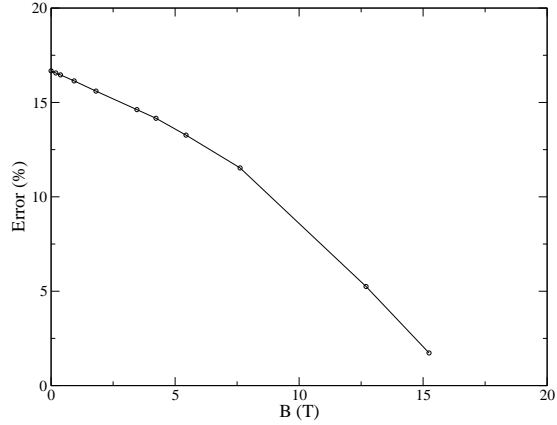}}
\end{center}
\end{figure} 
 In Table[1], we see that the error does not exceed $16.6667 \%$ for small values of B in the first energy level. It is important to note that the limit $m \to \infty$ is an extrapolation following the natural behavior of the results for great values of $m$. The lower acceptable value of $m$ for the computation of energies is $\frac{1}{2}$, because Eq.(\ref{eqn:34}) are no longer determinated. This implies that the high field limit, in the level $p=1$, occurs when $B$ is less than a critical value for which $m=\frac{1}{2}$. It is approximatively $B_{crit} = 19.053 T$. In Table [2], the error is less than $15 \%$ at the third level. Moreover, we remark that the greater the magnetic field is, the smaller is the error. In Fig.(3), we present the error as a function of $B$ in the first level, this shows us that the $R\to 0$ approximation is not restricted to the strong magnetic field limit, and we can use it in a more general scheme, where $R$ is small compared the average distance of the electron separation. It seems to be also good for weak $B$ since we have a limit in the error. Recalling that these results are made for WF with equals $m_{ij}$, the anyonic solutions are good solutions, and the WF in the limiting case $C=0$, with $m_{12}=m_{13}=m_{23}=m$, are also valid.
The analytic solutions, obtained in our paper, are appropriate to describe the system of three planar identical charged particles under perpendicular magnetic field, for all values of $B$ and not only in the high field limit.

\section{A new quantum Hall state?}
\emph{Unified form of wave functions for $2$ and $3$ particles system}

For this section, we must recall some results of [6], in particular the analytical solution of the two-body problem. Truong and Bazzali has shown that the WF of the relative motion is nothing else as:
\begin{eqnarray}
\phi_{2 rel} = \frac{e^{im\theta}}{\sqrt{2 \pi}} r^{|m|} \Pi_{p}(C_{2},r)\exp \left(-\frac{r^2}{4 l_{B}^{2}}\right)
\label{eqn:36}
\end{eqnarray}
where $r = |\vec{r_1} -\vec{r_2} |$ and $C_2$ is the ratio parameter in this case. If we include the center-of mass WF in the LLL, this equation reads in the real coordinates:
\begin{eqnarray}
\Psi_{2} =  ({\bf{r_{1}}} -{\bf{r_2}})^{m} \Pi_{p}(C_{2},|\vec{r_1} -\vec{r_2}|)\exp \left(-\sum_{i=1}^{2} \frac{r_{i}^{2}}{4 l_{B}^{2}}\right)
\label{eqn:37}
\end{eqnarray}
with the notation ${\bf{r}}^{m} =r^{|m|}e^{im\theta}$. If we compare this solution to the three-particle solution, we see that the main difference is $C_{3}=\frac{\sqrt{2}}{\sqrt{3}} C_{2}$. Moreover, using the relation $\sum_{i=1}^{3}r_{i}^{2} =\sum_{i=1}^{3}\xi_{i}^{2}$, we can rewrite our solution (\ref{eqn:09}) under the form:
\begin{eqnarray}
\Psi_{3} =  \prod_{i<j}^{k=3} ({\bf{r_i}} -{\bf{r_j}})^{m_{ij}} \Pi_{p}(C_{3},|\vec{r_i} -\vec{r_j}|) \exp \left({- \sum_{i=1}^{3} \frac{r_{i}^{2}}{4 l_{B}^{2}}}\right)
\label{eqn:38}
\end{eqnarray}

Let us transform these WF in complex coordinates, where $z = x+iy = r e^{i \theta}$. If we consider only states with all $m_{ij} = m$, i.e. anyons states and Landau states with equals angular momenta, we can rewrite WF (\ref{eqn:37}) and (\ref{eqn:38}) in an unified form:
\begin{eqnarray}
\Psi_{N_e} =  \prod_{i<j}^{k=N_e} (z_i -z_j)^{m} \Pi_{p}(C_{N_e},|z_i -z_j|)\exp \left(-\sum_{i=1}^{N_e} \frac{|z_{i}|^{2}}{4 l_{B}^{2}}\right)
\label{eqn:39}
\end{eqnarray}
where $N_e= 2, 3$ is the particle number. This WF have the following properties:
\begin{itemize}
\item it is exact for two particles and exact for three in the $R \to 0$ limit;
\item if $p$ is odd, it corresponds to a system of $N_e$ anyons which have a natural fractional statistics since $m$ is fixed by $C_{N_e}$;
\item if $p$ is even, it is also an anyon system when $C$ is finite, or a system of particles in extreme strong magnetic field if $C=0$;
\item have total angular momentum $M = m N_e (N_e -1)/2$.  
\end{itemize}
The question of the relevance of the Pauli principle does not arise here because we have an anyonic behavior for the quasi-particles. However, in the asymptotic case $C=0$, the system must verify the Pauli principle since we deal with standard particles (no fractional statistics). This demand is spontaneously verified when the $m_{ij}$ are odd. Note that all expressions are calculated for positive $m$, however the WF in the case $m<0$ can be expressed also easily, using the complex conjugate forms.
\\

\emph{Connection with the standard quantum Hall states}
\\
If we compare our WF(\ref{eqn:39}) to the so-called Laughlin WF:
\begin{eqnarray}
\Psi_{Laughlin} =  \prod_{i<j}^{j=N_e} (z_i -z_j)^{m} \exp \left(-\sum_{i=1}^{N_e} \frac{|z_{i}|^{2}}{4 l_{B}^{2}}\right)
\label{eqn:40}
\end{eqnarray}
we remark that it fits exactly for the asymptotic case $C=0$  in the lowest energy level $p=0$ (LLL). Effectively, since we have
\begin{eqnarray}
\Pi_{0}(0,|z_i -z_j|) = L_{0}^{m}(|z_i -z_j|^2) = 1,
\label{eqn:41}
\end{eqnarray}
 we can express our solution, for the $p^{th}$ Landau level as:
\begin{eqnarray}
\Psi_{N_e} =  \prod_{i<j}^{j=N_e}  \Pi_{p}(C_{N_e},|z_i -z_j|) \Psi_{Laughlin}.
\label{eqn:42}
\end{eqnarray}
Note that the coincidence occurs only for states with the same $m_{ij}$.

At this point, we can make several remarks. The Laughlin WF is effectively an analytical solution for $2$ particles and for $3$ in the infinite field limit and in the LLL. Our solution is richer than Laughlin solution, because it takes into account the higher Landau levels. Since the Laughlin WF deals with standard particles (no fractional statistics), the anyonic behavior in our model is brought by the biconfluent Heun (BCH) polynomials. Effectively, it is in the BCH theory that the second condition of quantization arises. As we have already said, when $p$ is odd, (\ref{eqn:42}) is a system of anyons with fractional statistics, and when $p$ is even, there is the two sets of solutions, anyonic and standard.

Since the WF (\ref{eqn:42}) is exact for $2$ particles and in the $R \to 0$ limit for $3$ particles, we propose the following  ansatz as a generalization for $N_e$ particles:
\begin{eqnarray}
\Psi =  \prod_{i<j}^{N_e}  \Pi_{p}(C,|z_i -z_j|) \Psi_{Laughlin}
\label{eqn:43}
\end{eqnarray}
 
This WF is a good one for the FQHE, and we hope that, for $p \neq 0$ and any value of $C$ ($B$), it corresponds to higher quantum Hall states. This question is postponed as a future work.

\section{Remarks and conclusion}

We have studied the case of three planar electrons under constant perpendicular magnetic field and with Coulomb repulsion. In the strong magnetic field limit, we have found analytic solutions of the Schr\"odinger equation using the Taut's variables. In this representation, each new quasi-particle can have two sets of behavior, anyons with fractional statistics for any $B$, or standard particles in Landau problem in the limiting case $C=0$. It's the main difference with Taut \cite{taut2}, who considered the standard quantization of the angular momentum which, with the second quantization condition, fixes the value of an additional parabolic scalar confinement potential of frequency $\omega_0$. In our picture, the solutions exist for all values of the external field and the angular momentum $m$ has now a continuous spectrum which is fixed by $C$. This anyonic behavior of these quasi-particles is possible in the quantum picture of relative motion of identical particles in two dimensions. In a natural way, the competition of the magnetic confinement and the Coulomb repulsion via the ratio parameter $C$ induces the fractional statistics which turns our quasi-particles into anyons.  

To justify our treatment, we have calculated the contribution of the dipole and the quadrupole in different states (different $p$, $m$ and $C$). We have proved that considering $R$ smaller than the average inter-particles' distance is also a good approximation outside the strong correlation limit since the error does not exceed $16 \%$. 

Finally, we have expressed in an unified way the WF of the $2$ particles system with the $3$ particles, and found that the Laughlin WF is a special case of ours. Effectivley, and with respect to the construction of the Laughlin states, the two WF coincide when $C=0$, i.e. when $B \to \infty$, and in the LLL. For other values of $p$, we have seen that the anyonic behavior is entirely brought by the BCH polynomials, but for the moment, we cannot relate it to some higher quantum Hall states. 

The Pauli principle is not in question in this problem since we deal with anyons which have fractional statistics. However, when $p$ is even, the Landau states exists for $C=0$ and the quasi-particles are standard electrons which do not have fractional statistics. But with $R \to 0$ and for $m$ odd (in the case where $m_{ij} = m$), Pauli principle is respected saving the quantum requirement in this case.

Finally, we have proposed a generalization to $N_e$ particles system which works for quantum Hall states when $C=0$ and in the LLL. The interesting aspect of this work is that the anyonic behavior is spontaneously generated by the competitive effects of the external forces. In a sense systems of planar charged particles in magnetic field develop  amazing effects. We reserve the study of the higher energy levels for future investigations, in particular the links with the CF states which have particular statistics.


\section{References}

\begin{thebibliography}{}
\bibitem{laughlin1}
R.B.Laughlin, Phys. Rev. Lett. \textbf{27}, (1983) 3383.
\bibitem{laughlin2}
R.B.Laughlin, Phys. Rev. Lett. \textbf{50}, (1983) 1395.
\bibitem{jain1}
J.K.Jain \& R.K.Kamilla, Phys. Rev. \textbf{B,52} 2798.
\bibitem{jain2}
J.K.Jain, Phys. Rev. \textbf{B,41} (1990) 7653.
\bibitem{jain3}
J.K.Jain, Phys. Rev. \textbf{B,51} (1995) 1752.
\bibitem{chakraborty}
T.Chakraborty \& P.Pietiläinen, \textit{The Fractional Quantum Hall Effect}, (Springer-Verlag, Berlin 1988).
\bibitem{prange}
R.E.Prange \& S.M.Girvin, \textit{The Quantum Hall Effect}, (Springer-Verlag, Berlin 1990).
\bibitem{taut1}
M.Taut, J. Phys. \textbf{A,27}, (1994) 1045.
\bibitem{khare}
A.Khare, \textit{Fractional Statistics and Quantum Theory}, (World Scientific, 1998).
\bibitem{vercin}
A.Ver\c{c}in, Phys. Lett.  \textbf{B,260}, (1991) 120.
\bibitem{truong}
T.T.Truong \& D.Bazzali, Phys. Lett. \textbf{A,269}, (2000) 186.
\bibitem{laidlaw}
M.G.G.Laidlaw \& C.M.DeWitt, Phys. Rev. \textbf{D,3}, (1975) 1375. 
\bibitem{leinaas}
J.M.Leinaas \& J.Myrheim, Nuovo Cimento \textbf{C,37}, (1977) 1.
\bibitem{goldin}
G.A.Goldin, R.Menikoff \& D.H.Sharp, J. Math. Phys. \textbf{22}, (1981) 1664.
\bibitem{khurana}
A.Khurana, Physics Today, (1989) 17.
\bibitem{wilczek}
F.Wilczek, \textit{Fractional Statistics and Anyon Superconductivity}, (World Scientific, 1989).
\bibitem{lerda}
A.Lerda, \textit{Anyons}, (Springer-Verlag, Berlin 1992).
\bibitem{taut2}
M.Taut, J. Phys. Cond. Math. \textbf{12}, (2000) 3689.
\bibitem{ballentine}
L.E.Ballentine, \textit{Quantum Mechanics: A Modern Development}, (World Scientific, 1998).
\bibitem{ronveaux}
A.Ronveaux, \textit{Heun's Differential Equation}, (Oxford University Press, 1995).
\bibitem{batola}
F.Batola,  \textit{Quelques propri\'et\'es de l'\'equation biconfluente de Heun, Th\`ese de troisi\`eme cycle} (Universit\'e Pierre et Maris Curie, Paris, 1977).
\bibitem{ralko}
A.Ralko \& T.T.Truong, Eur. Phys. J. \textbf{B,29} (2002) 335.
\end{thebibliography}
\end{document}